\documentclass[11pt]{article} 

\usepackage[utf8]{inputenc}
\usepackage[english]{babel}
\usepackage{graphicx}
\usepackage[a4paper,right=1in,left=1in,top=1in,bottom=1in]{geometry}
\usepackage[dvipsnames]{xcolor}
\usepackage[super]{natbib}
\usepackage{xcolor,colortbl}
\usepackage{tikz}
\usetikzlibrary{shapes,arrows,decorations.markings}
\usepackage{placeins}
\usepackage{pdflscape}

\usepackage{amsmath, amsfonts, amssymb} 
\usepackage[labelfont=bf]{caption}
\usepackage{setspace}
\usepackage{accents} 
\usepackage{color, soul}
\usepackage[title,titletoc,toc]{appendix}
\usepackage{rotating} 
\usepackage{float} 
\usepackage{multirow} 
\usepackage[position=top]{subfig}
\usepackage[subfigure]{tocloft}
\usepackage{tikz}
\usetikzlibrary{decorations.pathreplacing,calc}
\usepackage[many]{tcolorbox}
\newtcolorbox{rightbrace}{%
    enhanced jigsaw, 
    breakable, 
    frame hidden, 
    parbox=false,
}
\usepackage[explicit]{titlesec}
\graphicspath{ {images/} }

\usepackage{color}   
\usepackage{hyperref}

\title{Accounting for not-at-random missingness through imputation stacking}
\author{\textbf{Lauren J. Beesley$^{*1}$ and Jeremy M. G. Taylor$^{1}$} \\
$^{1}$University of Michigan, Department of Biostatistics\\
*Corresponding Author: lbeesley@umich.edu
}
\date{\today}

\begin{document}
\maketitle 

\allowdisplaybreaks
\raggedbottom

\begin{abstract}
Not-at-random missingness presents a challenge in addressing missing data in many health research applications. In this paper, we propose a new approach to account for not-at-random missingness \textit{after multiple imputation} through weighted analysis of stacked multiple imputations. The weights are easily calculated as a function of the imputed data and assumptions about the not-at-random missingness. We demonstrate through simulation that the proposed method has excellent performance when the missingness model is correctly specified. In practice, the missingness mechanism will not be known. We show how we can use our approach in a sensitivity analysis framework to evaluate the robustness of model inference to different assumptions about the missingness mechanism, and we provide R package \textit{StackImpute} to facilitate implementation as part of routine sensitivity analyses. We apply the proposed method to account for not-at-random missingness in human papillomavirus test results in a study of survival for patients diagnosed with oropharyngeal cancer.
\end{abstract}

Keywords: chained equations multiple imputation, fully conditional specification, not-at-random missingness, stacked imputation

\section{Introduction}
Multiple imputation is a popular and convenient strategy for addressing missing data in modern health research. One common strategy for obtaining imputations of the missing data involves filling in values for each variable with missingness one-by-one as part of an iterative algorithm. The problem of missing data handling then translates into assumptions about the imputation distribution used to fill in the missing values for each variable. This approach, called chained equations imputation or fully conditional specification, can be easily implemented using available statistical software. A robust statistical literature provides guidance for implementation in many common data analysis scenarios, but the majority of the statistical development and software rely on the key assumption that missingness is unrelated to unobserved data given observed data, called missing at random (MAR).  \\
\indent In many practical data settings, however, the restrictive MAR assumption may not hold. In the particular setting of health research, for example, results of medical tests may often be available for only a subset of patients. Symptom-informed medical testing can induce a relationship between whether a test is administered and the test result, even after adjusting for other observed data. When missingness relates to unobserved information, called missing not at random (MNAR), use of standard imputation strategies that rely on MAR assumptions can often result in biased inference.\cite{Little2002} \\
\indent Several researchers have developed modifications to the chained equations imputation procedure that incorporate corrections for MNAR missingness. Tompsett et al. (2018) proposed imputing each variable with missingness using a model that also adjusts for the missingness indicators for the \textit{other} variables.\cite{Tompsett2018} The relationship between missingness and the variable being imputed is incorporated through a fixed offset in the imputation model with a corresponding sensitivity parameter. The random indicator method proposed in Jolani (2012) avoids the use of fixed sensitivity parameters, but existing software implicitly assumes a logistic regression model for missingness with main effects only, which may be violated in practice. Both methods rely on some degree of approximation in the distribution of missing values given observed values.\cite{Jolani2012} \\
\indent Additionally, these approaches to handling MNAR missingness assume that the analyst fitting the target model is also the analyst imputing the data. Carpenter et al. (2007) provides a sensitivity analysis approach that re-weights parameter estimates from multiple imputations generated under MAR assumptions, with the structure of the weights used to account for the MNAR missingness.\cite{Carpenter2007} This method can perform well in settings where the MNAR missingness is weak, but several authors have noted that this method can perform poorly when MNAR missingness is strong.\cite{Rezvan2015} Corrections to this method provided in Smuk (2015) only partially address this issue.\cite{Smuk2015} \\ 
\indent In this paper, we propose a new approach for addressing not-at-random missingness that takes advantage of recent advances in the area of stacked multiple imputations.\cite{Beesley2020} In the proposed approach, multiple imputations obtained under MAR assumptions are stacked and augmented with a weight related to the assumed MNAR missingness mechanism. When all models are correctly specified, we can obtain valid estimates of parameters of interest (e.g. means, regression model parameters, etc.) by performing a weighted version of the target analysis on the stacked multiple imputations. We describe several strategies for obtaining corresponding standard errors based on previous work by Beesley and Taylor (2020) and Bernhardt (2019).\cite{Beesley2020,Bernhardt2019} We demonstrate through simulation that the proposed method has excellent performance when the missingness model is correctly specified. In practice, the missingness mechanism will not be known. We show how we can use our approach in a sensitivity analysis framework to evaluate the robustness of model inference to different MNAR assumptions. We apply the proposed method to account for potential MNAR missingness in human papillomavirus (HPV) test results in a study of survival for patients diagnosed with oropharyngeal cancer. We also provide R package \textit{StackImpute} to facilitate implementation as part of routine sensitivity analyses to deviations from MAR.

\section{Imputation stacking approach for single variable MNAR missingness}

\subsection{Notation and assumptions}
Let $Z$ be a $n \times p$ matrix containing $p$ variables measured on $n$ independent subjects such that the first $k$ variables in $Z$ are missing for some subjects and the last $p-k$ variables (denoted $W$) are fully-observed for all subjects. Let $R_{ij}$ be an indicator for whether variable $j$ is measured for subject $i$ in the data. Our goal is to obtain multiple imputations of the missing values in $Z$, with which we will perform some target analysis. For example, we may be interested in the mean of the $j^{th}$ variable in $Z$ or a regression model of the first variable on the others. Throughout, let $Z_{i.}$ denote data for the $i^{th}$ subject, and let $Z_{.j}$ denote the $j^{th}$ variable. Define rows and columns in matrices $R$ and $W$ similarly. Let $Z_{i,-j}$ denote the elements in $Z_{i.}$ excluding the $j^{th}$ variable. We will assume data are independent across subjects. \\
\indent We imagine the missingness pattern $R_{ij}$ observed in the data is a \textit{data realization} of a corresponding \textit{random variable}, denoted $\mathcal{R}_{ij}$. Collectively, we call these random indicators $\mathcal{R}$. Under missing at random (MAR) assumptions, the joint distribution of $\mathcal{R}$ may depend only on fully-observed variables $W$, and the mechanism generating data missingness can be ignored during data imputation.\cite{Little2002} However, it is possible that missingness depends on unobserved information in $Z$, called missing not at random (MNAR). Here, we consider a particular generalization of the MAR setting where the first variable in $Z$, denoted $Z_{.1}$, may be MNAR. For an extension of these methods under multiple variable MNAR missingness, see \textbf{Supplementary Section A}. \\
\indent Suppose we partition the joint model for missingness as 
 \begin{align*}
 f(\mathcal{R}_{i1}, \hdots,  \mathcal{R}_{ik} \vert Z_{i.}) = f(\mathcal{R}_{i2}, \hdots,  \mathcal{R}_{ik} \vert Z_{i.}, \mathcal{R}_{i1}) f(\mathcal{R}_{i1} \vert Z_{i.}).
 \end{align*}
where $f$ denotes the distribution function for the corresponding variables. We will assume the following:
\begin{enumerate}
  \item $Z_{i2}, \hdots, Z_{ik}$ are MAR, with $f(\mathcal{R}_{i2}, \hdots,  \mathcal{R}_{ik} \vert Z_{i.}, \mathcal{R}_{i1}) = f(\mathcal{R}_{i2}, \hdots,  \mathcal{R}_{ik} \vert W_{i.})$. 
  \item $Z_{i1}$ may be MNAR, with $f(\mathcal{R}_{i1} \vert Z_{i.}) = f(\mathcal{R}_{i1} \vert Z_{i1}, W_{i.})$.
\end{enumerate}
In \textit{Assumption 2}, we allow $Z_{i1}$ to be MNAR such that its missingness depends on the true value of $Z_{i1}$ but does not depend on the other variables with missing values.

\subsection{Imputation and importance sampling}  \label{imputation}
\indent Let $Z_{i,mis}$ and $Z_{i,obs}$ denote the missing and observed elements of $Z_{i.}$, respectively. Under a full joint model for the variables with missingness, we can impute missing values of $Z_{i.}$ from $f(Z_{i,mis} \vert Z_{i,obs},\mathcal{R}_{i.}=R_{i.})$. In practice, we often approximate a draw from the full joint distribution by iteratively drawing missing values for each variable in $Z$ from its full conditional distribution, $f(Z_{ij} \vert Z_{i,-j},\mathcal{R}_{i.}=R_{i.})$. Then, we repeat this iterative process many times to obtain $M$ imputed datasets. Rather than specifying the full joint model for all variables with missingness, a chained equations strategy involves directly specifying a model for each full conditional distribution, $f(Z_{ij} \vert Z_{i,-j},\mathcal{R}_{i.}=R_{i.})$. It can be challenging in general to determine how to specify these conditional models as a function of $R_{i.}$. Under \textit{Assumptions 1-2}, however, these imputation distributions can be simplified. \\
\indent First, we consider imputation for $Z_{ij}$ in $Z_{i2}, \hdots, Z_{ik}$. Under \textit{Assumptions 1-2},  $f(Z_{ij} \vert Z_{i,-j},\mathcal{R}_{i.}=R_{i.}) = f(Z_{ij} \vert Z_{i,-j}, \mathcal{R}_{ij}=1) =  f(Z_{ij} \vert Z_{i,-j})$. This is the same distribution we would use to impute $Z_{ij}$ under standard MAR assumptions, and we can apply our usual strategies for performing this imputation, e.g. by approximating $f(Z_{ij} \vert Z_{i,-j})$ with a regression model.  \\
\indent Now, we consider the imputation distribution for $Z_{i1}$. Under \textit{Assumption 1}, we can impute missing $Z_{i1}$ from 
 \begin{align*}
f(Z_{i1} \vert Z_{i,-1},\mathcal{R}_{i.}=R_{i.}) = f(Z_{i1} \vert Z_{i,-1},\mathcal{R}_{i1}=0)
\end{align*}
Parameters from this distribution are not identified from the observed data without additional assumptions. However, we note that 
 \begin{align*}
&f(Z_{i1} \vert Z_{i,-1}, \mathcal{R}_{i1} = 0) = \frac{P(\mathcal{R}_{i1} = 0 \vert Z_i)}{P(\mathcal{R}_{i1} = 0 \vert Z_{i,-1})} f(Z_{i1} \vert Z_{i,-1})  \\
&= \frac{P(\mathcal{R}_{i1} = 0 \vert Z_i)}{P(\mathcal{R}_1 = 0 \vert Z_{i,-1})} \left[ \sum_r f(Z_{i1} \vert Z_{i,-1}, \mathcal{R}_{i1} = r) P(\mathcal{R}_{i1} = r \vert Z_{i,-1})  \right] \nonumber\\
&= P(\mathcal{R}_{i1} = 0 \vert Z_i) \left[ f(Z_{i1} \vert Z_{i,-1}, \mathcal{R}_{i1} = 1) \frac{P(\mathcal{R}_{i1} = 1 \vert Z_{i,-1})}{P(\mathcal{R}_{i1} = 0 \vert Z_{i,-1})} + f(Z_{i1} \vert Z_{i,-1}, \mathcal{R}_{i1} = 0)   \right] \nonumber
\end{align*}
The term $f(Z_{i1} \vert Z_{i,-1}, \mathcal{R}_{i1} = 0)$ appears on both the left and the right of this expression. Rearranging this expression and using that $P(\mathcal{R}_{i1} = 0 \vert Z_{i.})=P(\mathcal{R}_{i1} = 0 \vert Z_{i1}, W_{i.})$, we have that 
 \begin{align} \label{imputeZ1}
&f(Z_{i1} \vert Z_{i,-1},  \mathcal{R}_{i1} = 0) \propto  \frac{P(\mathcal{R}_{i1} = 0 \vert Z_{i1}, W_{i.})}{1-P(\mathcal{R}_{i1} = 0 \vert Z_{i1}, W_{i.})} f(Z_{i1} \vert Z_{i,-1},  \mathcal{R}_{i1} = 1)
\end{align}
Unlike $f(Z_{i1} \vert Z_{i,-1}, \mathcal{R}_{i1} = 0)$, there is information in the data to estimate parameters in distribution $f(Z_{i1} \vert Z_{i,-1},  \mathcal{R}_{i1} = 1)$. The term $P(\mathcal{R}_{i1} = 0 \vert Z_{i1}, W_{i.})$ is not identified from the observed data, and we will need to make untestable assumptions about this distribution. We will address this challenge later on. Even if this missingness probability were known, drawing from \ref{imputeZ1} can still be difficult, since the distribution may only be known up to proportionality in some cases. Rejection sampling and other statistical techniques can be applied to draw from \ref{imputeZ1} directly, but these approaches can be computationally expensive and require custom software.  \\
\indent One option is to approximate a draw from \ref{imputeZ1} using importance sampling as in Tanner (1993) and Little and Rubin (2002).\cite{Tanner1993,Little2002} Define functions $j(z) = f(z \vert Z_{i,-1},  \mathcal{R}_{i1} = 1)$ and $h(z) = f(z \vert Z_{i,-1},  \mathcal{R}_{i1} = 0)$. We can approximate a draw from $h(z)$ by drawing multiple candidate imputations $z^1, \hdots, z^M$ from $j(z)$. Then, we select candidate draw $z^m$ with probability proportional to $h(z^m)/j(z^m)$ to obtain a \textit{single} imputation of $Z_{i1}$. This process can then be repeated multiple times to obtain multiple imputations. This importance sampling method can be applied if 
\begin{enumerate}
  \item The support of $j(z)$ contains the support for $h(z)$.
  \item Function $h(z)/j(z)$ is bounded. 
\end{enumerate}
The first requirement may often be met for $j(z)$ and $h(z)$ as defined above. However, many candidate draws may be needed when $j(z)$ and $h(z)$ are very different, i.e. the distribution of observed $Z_{i1}$ is very different than the distribution of missing $Z_{i1}$. The second requirement is satisfied if $w(z) = \frac{P(\mathcal{R}_{i1} = 0 \vert z, W_{i.})}{P(\mathcal{R}_{i1} = 1 \vert z, W_{i.})}$ is bounded in $z$. $w(z)$ will, of course, be bounded below by 0. However, additional assumptions are needed to ensure $w(z)$ is bounded above. We can ensure that $w(z)$ is bounded above if we assume there is some (possibly small) probability $\epsilon$ such that $\epsilon<P(\mathcal{R}_{i1} = 1 \vert z, W_{i.})$ for all $z$. In other words, the probability of observing $Z_{i1}$ must always be non-zero. While this may not strictly hold for some $Z_{i1}$ (e.g. those defined on the real line under logistic regression), we may still reasonably apply this importance sampling strategy if the probability of drawing very extreme candidates for $Z_{i1}$ is small.    \\
\indent The above approach can become computationally expensive, since we need many candidate draws from $j(z)$ in order to obtain a single imputed value from $h(z)$. An alternative approach is to first obtain $M$ multiple imputations of $Z_{i1}$ and to weight these multiple imputations proportional to $w(z)$ in the data analysis. The exact way in which these weights should be carried through in the analysis of multiply imputed data, however, is not obvious. Previously, Carpenter et al. (2007) proposed a strategy for incorporating such weights into analysis of multiply imputed $Z_{.1}$ as described in \textbf{Section \ref{carpenter}} below.\cite{Carpenter2007} In this paper, we will propose a different strategy to incorporate such weights into data analysis that maintains the simplicity of the method in Carpenter et al. (2007) but gives better properties in terms of bias in estimating parameters of interest.\cite{Carpenter2007}

\subsection{Weighting strategy of Carpenter et al. (2007)} \label{carpenter}
Suppose we obtain multiple imputations of missing values in $Z_{i1}$ \textit{as if missingness were MAR} from $f(Z_{i1} \vert Z_{i,-1},  \mathcal{R}_{i1} = 1)$. Let $\theta$ denote our parameter of interest. A common strategy for obtaining the final estimate of $\theta$ for multiply imputed data is to take the average of parameter estimates obtained for each of the individual imputations, here denoted $\hat \theta_1, \hdots, \hat \theta_M$. To account for the MNAR missingness, Carpenter et al. (2007) proposes taking a \textit{weighted average} of these estimates.\cite{Carpenter2007} The structure of the weight proposed by Carpenter et al. (2007) was motivated by the relation in \ref{imputeZ1}, in the special case where the missingness model for $Z_{i1}$ can be approximated by the following logistic regression: $\text{logit}\left( P(\mathcal{R}_{i1} = 1 \vert Z_{i1}, W_{i.})\right) = \phi_0 + \phi_1 Z_{i1} + \phi_W^T W_{i.}$. Omitting some details, the weight for imputation $m$ proposed in Carpenter et al. (2007) is defined as
 \begin{align} \label{carpenterweights}
\alpha_{m} \propto \text{exp}\left( - \phi_1 \sum_{i: R_{i1} = 0}^n Z_{i1m} \right)
\end{align}
where the $\alpha$'s are rescaled so that $\sum_{k=1}^M \alpha_k = 1$ and $Z_{i1m}$ denotes the $m^{th}$ imputation of $Z_{i1m}$. Point estimates and standard errors under MNAR are then obtained as follows: 
\begin{align} \label{carpenterstderrs}
\hat \theta_{MNAR} &= \sum_{m=1}^M \alpha_{m} \hat \theta_m\\
Var(\hat \theta_{MNAR}) &= \sum_{m=1}^M \alpha_m Var(\hat \theta_m)  + \left( 1+1/M\right)  \sum_{m=1}^M \alpha_m \left[\hat \theta_m - \hat \theta_{MNAR}\right]^2 \nonumber
\end{align}
In this analysis, $\phi_1$ is treated as a sensitivity parameter, and the final analysis is performed multiple times across a plausible range of $\phi_1$ values. This analysis approach is easy to implement and allows the imputation to be separated from the handling of MNAR missingness. Unlike more commonly-used MNAR sensitivity analysis strategies such as those in Tompsett et al. (2018), this approach does not require $Z_{.1}$ to be imputed separately for each fixed value of the sensitivity parameter.\cite{Tompsett2018}  \\
\indent As discussed in Carpenter et al. (2013), however, this approach requires the true $\theta$ value to be within the range of the $\hat \theta_j$ estimates obtained from each of the imputed datasets under MAR.\cite{Carpenter2013} \cite{Rezvan2015} demonstrates that the approach in Carpenter et al. (2007) can produce substantial bias when this assumption is not met.\cite{Carpenter2007} Additionally, Rezvan et al. (2015) shows that this approach does not guarantee a consistent estimate of $\theta$ even when this assumption is met.\cite{Rezvan2015} Smuk (2015) provides a correction to these weights that may reduce this bias in some cases, but this correction can only be applied when we have MNAR missingness in a single, normally-distributed variable.\cite{Smuk2015} Additionally, the method in Smuk (2015) cannot provide consistent parameter estimates when the true value of $\theta$ is outside the range of estimates obtained under MAR assumptions.

\subsection{Proposed weighting and analysis strategy}
\begin{figure}[htbp!]
  \centering
\caption{Visualization of proposed imputation and stacked data analysis procedure $^1$}
\includegraphics[trim={0cm 0cm 0cm 0cm}, clip, width=6in]{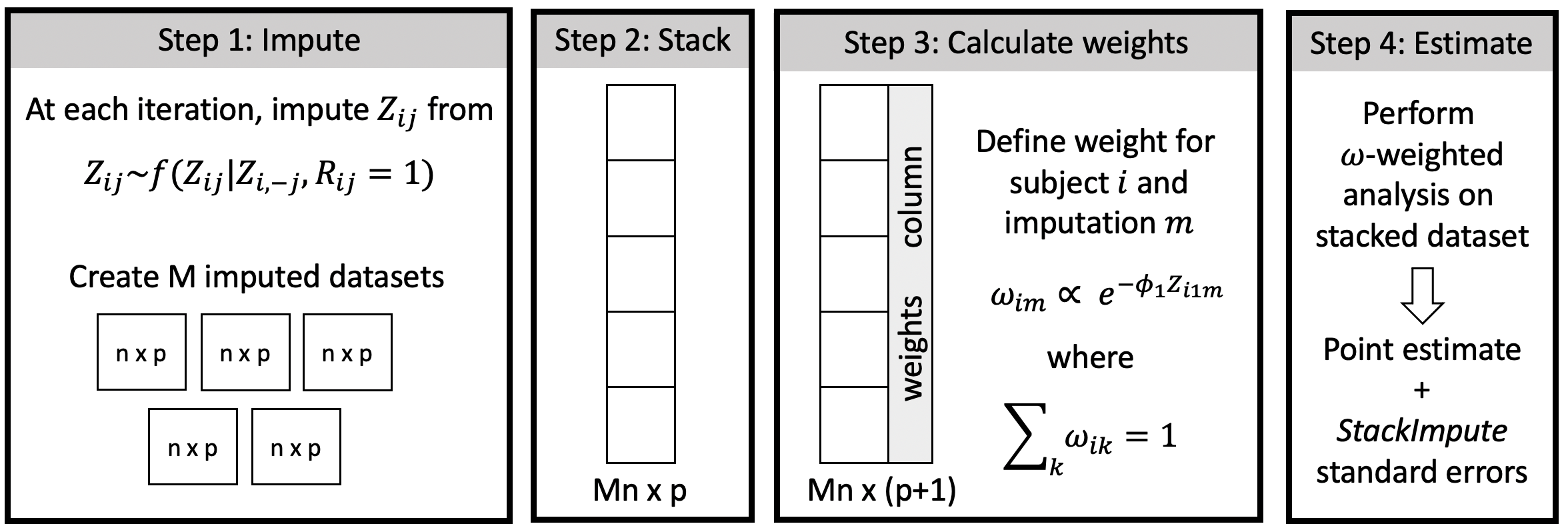}
\caption*{ \footnotesize  $^1$ $\phi_1$ is a sensitivity analysis parameter. Steps 3-4 can be repeated for multiple values of $\phi_1$. R \textit{StackImpute} can be used to estimate standard errors using the methods in \textbf{Section \ref{varestimation}}. 
}
\label{methoddiagram}
\end{figure}

\indent The importance sampling logic in \textbf{Section \ref{imputation}} implies that each imputed value of $Z_{i1}$ should be weighted proportional to $\frac{P(\mathcal{R}_{i1} = 0 \vert Z_{i1}, W_{i.})}{1-P(\mathcal{R}_{i1} = 0 \vert Z_{i1}, W_{i.})} $, where weights are rescaled to sum to 1 \textit{for each subject}. Instead, the Carpenter et al. (2007) approach weights vector $Z_{.1}$ by the \textit{product} of the unscaled weights for each individual $Z_{i1}$, and the resulting aggregate weights are scaled to sum to 1 across imputations.\cite{Carpenter2007} This approach no longer distinguishes between ``good" and ``bad" imputations of individual $Z_{i1}$ (in terms of their corresponding weights relative to the target distribution under MNAR) and instead considers imputed datasets in aggregate. We posit that this weight aggregation step is the primary source of residual downstream bias in the final analysis.  \\
\indent To address this issue, we propose maintaining separate weights at the individual level and instead performing our analysis using imputation stacking. The idea behind imputation stacking is that multiply imputed datasets are stacked on top of each other to form a large, $Mn$ by $p$ matrix. $\theta$ can then be estimated by performing our target analysis on the stacked dataset. Previous work has shown that this approach can produce estimates of $\theta$ equivalent to analysis by Rubin's combining rules.\cite{Wang1998} Historically, this imputation stacking approach has been difficult to implement owing to the lack of easy-to-use estimators for corresponding standard errors. Wood et al. (2008) proposed a simple method for estimating standard errors for stacked data analysis, but we showed that this approach can result in substantially biased standard error estimates in many settings.\cite{Wood2008,Beesley2020} Recently, Beesley and Taylor (2020) proposed a new strategy for estimating valid standard errors based on the observed data information principle of Louis (1982).\cite{Beesley2020,Louis1982} An alternative bootstrap-based estimator has also been proposed in Bernhardt (2019).\cite{Bernhardt2019} These advancements have made the stacked imputation strategy an accessible and appealing analysis framework. \\
\indent We propose handling MNAR missingness in $Z_{i1}$ through a weighted analysis of the stacked data as follows. This approach is summarized in \textbf{Figure \ref{methoddiagram}}. \\
$\bullet$ \textbf{Step 1:} Obtain $M$ multiple imputations of $Z$ assuming ignorable missingness (MAR).\\
Obtain $M$ multiple imputations of the missing data using chained equations imputation, where each variable $Z_{ij}$ with missingness is imputed from a regression model approximation to $f(Z_{ij} \vert Z_{i,-j}, \mathcal{R}_{ij} = 1)$, denoted $\tilde f$. Following logic commonly-used in chained equations imputation, we approximate a draw from this distribution by first drawing the corresponding parameter, which we will denote $\beta_j$, as follows:
\begin{align} \label{chainedeqns}
\text{Draw parameter $\beta_j$ from }& \tilde f(\beta_j \vert Z_{.,-j}^{imp}, Z_{.j}^{obs}, \mathcal{R}_{.j} = 1)\\
\text{Impute missing $Z_{ij}$ from }& \tilde f(Z_{ij} \vert Z_{i,-j},  \mathcal{R}_{ij} = 1; \beta_j) \nonumber
\end{align}
where $Z_{.,-j}^{imp}$ denotes the most recent imputed version of variable $Z_{.,-j}$ and $Z_{.j}^{obs}$ denotes the fully-observed elements in $Z_{.j}$. In practice, we can obtain a draw of $\beta_j$ from a multivariate normal distribution with mean and covariance matrix estimated from fitting a model for $Z_{.j} \vert Z_{.,-j}$ to the subset of recently imputed data  \textit{for which $Z_{.j}$ is observed}. Alternatively, $\beta_j$ could be obtained by fitting a model to a bootstrap sample of the same subset of the imputed data. The key here is that parameters used for imputing $Z_{.1}$ are drawn only using data from subjects with \textit{observed} $Z_{i1}$, since the distribution of $Z_{i1}$ given $R_{i1} = 1$ is not the same as the unconditional distribution. Fortunately, this approach for obtaining parameter draws is used by many commonly-used statistical packages to impute missing data under MAR assumptions, e.g. package \textit{mice} in R and \textit{PROC MI} in SAS. Therefore, we can often impute $Z$ using standard imputation software assuming ignorable missingness. \\
$\bullet$ \textbf{Step 2:} Stack multiple imputations\\
Generate a $Mn$ by $p$ dataset by stacking the $M$ multiple imputations on top of each other. \\
$\bullet$ \textbf{Step 3:} Calculate weights\\
Let $Z_{i1m}$ denote the $m^{th}$ multiple imputation of $Z_{i1}$. We then define weights 
 \begin{align} \label{weightsZ1}
& \omega_{im} \propto \frac{P(\mathcal{R}_{i1} = 0 \vert Z_{i1m}, W_{i.})}{1-P(\mathcal{R}_{i1} = 0 \vert Z_{i1m}, W_{i.})}  
\end{align}
that are rescaled such that $\sum_{k=1}^M \omega_{ik} = 1$. We augment each row of the stacked dataset with the corresponding weight, where rows corresponding to subjects with observed $Z_{i1}$ are assigned weight $1/M$. \\
$\bullet$ \textbf{Step 4:} Estimate parameter of interest\\
Let $\theta$ be our parameter of interest. Estimate $\hat \theta$ by performing a \textit{weighted} version of our target analysis to the stacked dataset, where weights are defined as in Step 3. For example, if our goal is to estimate the mean of $Z_{.1}$, we can estimate a weighted mean from the stacked dataset. If all models are correctly specified, this will produce a valid estimate for $\theta$. We can estimate corresponding standard errors using the strategies in \textbf{Section \ref{varestimation}}. We provide software for applying these standard error estimators in R package \textit{StackImpute}.

\subsection{Modeling missingness}
The structure of the weights in \ref{weightsZ1} depends on an assumed model for whether or not $Z_{i1}$ is observed, and a key limitation of this approach is that this missingness relationship may often be unknown. Suppose, however, that we posit a regression \textit{model structure} for this missingness model as follows: 
 \begin{align} \label{linkZ1}
&g\left( P(\mathcal{R}_{i1} = 1 \vert Z_{i1}, W_{i.})\right) = \phi_0 + \phi_1 Z_{i1} + \phi_2^T W_{i.}
\end{align}
where $\phi_W$ may be a vector. Then, we can define 
 \begin{align} \label{weights_link}
& \omega_{im} \propto \frac{1-g^{-1}\left( \phi_0 + \phi_1 Z_{i1m} + \phi_2^T W_{i.} \right)}{g^{-1}\left( \phi_0 + \phi_1 Z_{i1m} + \phi_2^T W_{i.} \right)}.
\end{align}
In general, this weight will not have a nice form, and it depends on unknown parameters $\phi_0$, $\phi_1$, and $\phi_2$. Generally, $\phi_1$ will not be identifiable from the observed data, and we will instead treat it as a fixed sensitivity parameter as is done in Carpenter et al. (2007).\cite{Carpenter2007} For a fixed value of $\phi_1$, parameters $\phi_0$ and $\phi_2$ can be estimated by fitting the model in \ref{linkZ1} with fixed offset $\phi_1 Z_{i1m}$ to the (unweighted) dataset obtained by stacking the multiple imputations. This can be repeated to generate weights for different fixed values of $\phi_1$ within a plausible window. Unlike usual sensitivity analysis strategies applied within the chained equations algorithm as in Tompsett et al. (2018), this sensitivity parameter can be directly interpreted as the variable's association with its own missingness.\cite{Tompsett2018} \\
\indent One downside of this strategy is that it requires us to specify the functional relationship between missingness in $Z_{.1}$ and fully-observed variables, $W_{.1}$. In the special case where \ref{linkZ1} corresponds to a logistic regression, however, the structure of the weights simplifies as follows: 
 \begin{align} \label{logistic}
& \omega_{im} \propto \text{exp}\left(-\phi_1 Z_{i1m}\right)
\end{align}
The contribution of $W_{.1}$ drops out of these weights after we rescale the weights such that $\sum_{k=1}^M \omega_{im} = 1$. In other words, the weights become a simple function of (1) the multiple imputations of $Z_{i1}$ and (2) the fixed sensitivity parameter, $\phi_1$. This result holds true for a more general class of logistic regression missingness models where $W_{i.}$ is allowed to have more complicated relationships with $\mathcal{R}_{i1}$, including non-linear effects or interactions between variables in $W_{i.}$.

\section{Variance estimation strategies} \label{varestimation}
In Beesley and Taylor (2020), we proposed a strategy for estimating standard errors for $\hat \theta$ obtained using maximum likelihood estimation based on stacked and weighted data as follows.\cite{Beesley2020} Let $J_{com}$ be the negative of the second derivative matrix of the complete data log-likelihood function for the target analysis, and let $U_{com}$ be the first derivative matrix of the complete data log-likelihood function. Let $J^i_{com}(\theta)$ and $ U^i_{com}(\theta)$ be the contributions to the complete data information matrix and score matrix for subject $i$ respectively. We approximate
\begin{align}\label{louis}
I_{obs}(\theta) &\approx \sum_i  \sum_m \omega_{im} J^i_{com}(Z_{i.m}; \theta)  -  \sum_i \sum_m \omega_{im}  \left[U^i_{com}(Z_{i.m}; \theta)  -   \bar U^i_{com}(Z_{i..}; \theta)  \right]^{\otimes 2}
\end{align}
where $\bar U^i_{com}(Z_{i..}; \theta) = \sum_k \omega_{ik}  U^i_{com}(Z_{i.k}; \theta)$. We can evaluate this expression at the maximum likelihood estimator for $\theta$, $\hat \theta$, obtained from maximizing the complete data log-likelihood using the  weighted, stacked dataset. Inverting the resulting matrix $I_{obs}(\hat \theta)$ will provide an estimate for the observed data covariance matrix for $\hat \theta$. \\
\indent A limitation of the estimator in \ref{louis} is that it requires us to obtain the complete data score and information matrices and can only be applied when our target analysis is maximum likelihood estimation with a valid log-likelihood function. Additionally, this approach can produce inaccurate or even negative variances when $n$ is small (e.g., n = 100). An alternative method proposed in Bernhardt (2019) uses bootstrap methods to account for so-called ``between imputation" variation as follows.\cite{Bernhardt2019} Let $V_{stack}$ be the estimated covariance matrix output by the stacked and weighted analysis, obtained using standard error estimation strategies that account for the weights. For example, for a generalized linear model with a dispersion parameter, the dispersion parameter must also be estimated using weighted residuals. The matrix $V_{stack}$ represents the ``within imputation" variation. To capture the ``between imputation" variation, we estimate $\theta$ on many bootstrap replicates of the stacked data. Unlike standard bootstrap replication, we obtain each bootstrap replicate of the stacked data by drawing with replacement from the set of indices $\{1, \hdots, M\}$ corresponding to the $M$ imputed datasets and then construct the bootstrapped stacked dataset composed of the drawn $M$ imputed datasets, where individual imputed datasets may appear in the stack multiple times. We then re-scale weights $\omega_{im}$ in the bootstrapped stack so that the weights again sum to 1 within individuals. Let $V_{between}$ be the estimated covariance matrix of the resulting $\hat \theta$ estimates across bootstrap samples of the imputed datasets. We then estimate the overall covariance matrix as follows: 
 \begin{align} \label{between}
Var(\hat \theta_{MNAR}) &=  V_{stack} + \left( 1+M\right) V_{between} 
\end{align}
One unappealing feature of the bootstrap-based estimator for $V_{between}$ proposed in Bernhardt (2019) is that it may require a large number of bootstrap samples, which can result in slow estimation.\cite{Bernhardt2019} Instead, we propose estimating the ``between imputation" variation using a jackknife estimator, defined with respect to leave-one-out \textit{imputations}. We estimate $V_{between}$ as
 \begin{align} \label{jackknife}
& V_{between} = \frac{M-1}{M} \sum_m \left[\hat \theta^{(m)} - \bar \theta \right]^{\bigotimes 2} & \bar \theta = \frac{1}{M} \sum_m \hat \theta^{(m)}  \nonumber
\end{align}
where $\hat \theta^{(m)}$ is estimated by fitting the model on the stacked data excluding the $m^{th}$ multiple imputation, again re-scaling the weights to sum to 1 within subjects.

\section{Simulations reproducing Rezvan et al. (2015): MNAR outcome missingness}  \label{rezvansimssec}
\indent Rezvan et al. (2015) conducted a simulation study to demonstrate settings in which the method from Carpenter et al. (2007) does and does not perform well.\cite{Rezvan2015,Carpenter2007} Here, we reproduce this simulation study and compare the results obtained using our proposed method to complete case analysis and the method in Carpenter et al. (2007).

\subsection{Simulation set-up}
\indent Suppose our goal is to estimate the association between outcome variable $Z_{1}$ (partially missing) and covariate $Z_{2}$ (fully-observed), and suppose we have MNAR missingness in outcome $Z_{1}$ dependent on the true value of $Z_{1}$ and $Z_{2}$. For each of several simulation settings, we generate 1000 simulated datasets of n = 100 or 1000 subjects. In all settings, we generate $ Z_{2} \sim N(0,1)$. We then generate $Z_1$ using one of two models: (1) $Z_{1} \sim N(0.5 Z_{2},1)$ or (2) $\text{logit}(P(Z_{1}=1  \vert Z_{2})) = 0.5 Z_{2}$. In both settings, we then impose roughly 50\% missingness in $Z_1$ using the following logistic regression model:  $\text{logit}(P(Z_{1} \text{ observed } \vert Z)) = \phi_1 Z_{1} + Z_{2} $ where $\phi_1 = 1$, 0.5, or 0. We then obtain $M$ multiple imputation of the missing values of $Z_1$ under MAR assumptions, where $M$ takes values 5, 10, 50, 100, 500, or 1000. We apply the proposed method and the method in Carpenter et al. (2007) to estimate parameters in the outcome model (either linear or logistic regression for $Z_1 \vert Z_2$).\cite{Carpenter2007} We obtain parameter estimates under different \textit{assumed} values for $\phi_1$, including 0, 0.2, 0.5, 0.8, 1, and 1.2. 

\subsection{Point estimation }
\begin{figure}[htbp!]
  \centering
\caption{Average estimated outcome model parameters across 1000 simulated datasets under MNAR missingness in outcome $Z_1$.}
\subfloat[][Estimates assuming true $\phi_1$ is known (results shown for true $\phi_1 = 1$)]{\includegraphics[trim={0cm 0cm 0cm 0cm}, clip, width=6in]{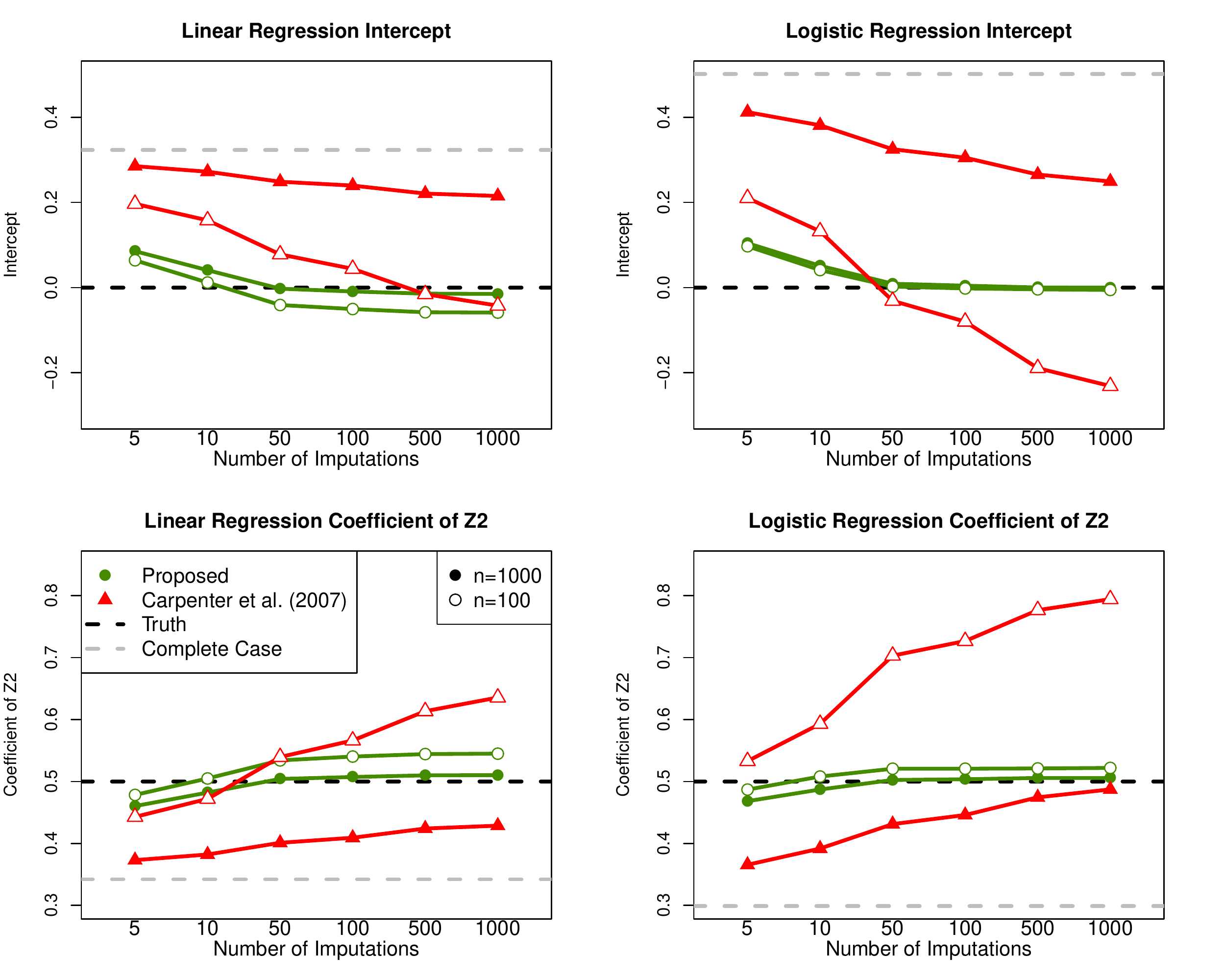}}\\
\vspace{0.5cm}
\subfloat[][Estimates across assumed values for $\phi_1$. Results shown true $\phi_1 = 1$ (left) and 0.5 (right)$^1$ ]{\includegraphics[trim={0cm 0cm 0cm 0cm}, clip, width=6in]{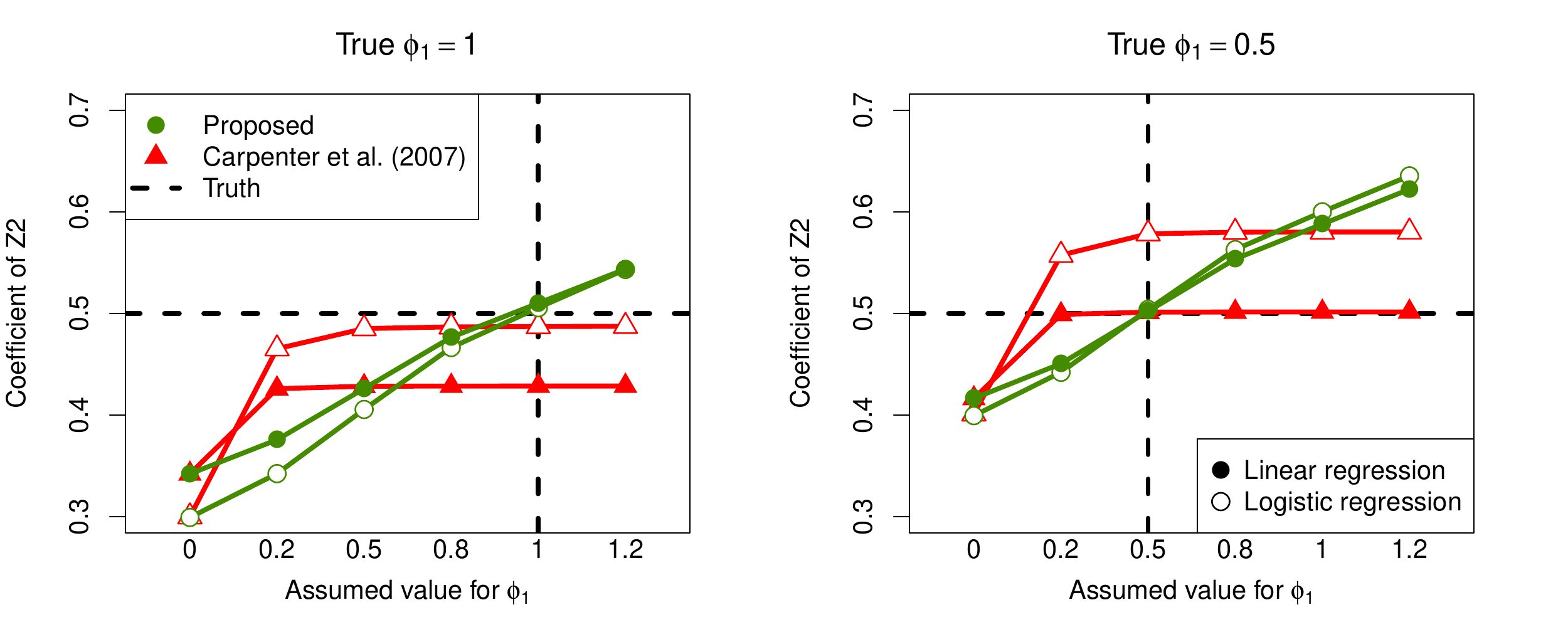}}
\caption*{ \footnotesize $^1$ Results shown for $M=1000$ and $n=1000$. }
\label{rezvan}
\end{figure}

\indent \textbf{Figure \ref{rezvan}a} shows the average estimated values for outcome model parameters across 1000 simulated datasets, assuming $\phi_1$ is known to equal 1. Complete case analysis and analysis of imputations obtained under MAR (not shown, similar to complete case estimates) produced substantial bias in estimating outcome model parameters. Although it results in reduced bias compared to complete case analysis, the method in Carpenter et al. (2007) produced substantial residual bias even when $n$ and $M$ were large.\cite{Carpenter2007} In contrast, the proposed method had small or negligible bias in estimating all model parameters as long as $M$, the number of multiple imputations, was large enough (e.g. $\geq 50$). For example, the method in Carpenter et al. (2007) resulted in biases in the linear regression coefficient of $Z_2$ up to 21\% for n=100 and up to 14\% for n=1000 with $M = 1000$. In comparison, the proposed method gave much smaller biases, with corresponding biases in the linear regression coefficient of $Z_2$ down to 9\% for n=100 and down to 2\% for n=1000. Results were similar for $\phi_1 = 0.5$. In \textbf{Supplementary Section C}, we describe a second simulation study in which missingness was generated for multiple covariates, one of which was MNAR. Results were similar.\\
\indent We also compare the performance of the proposed method in terms of point estimation with two existing MNAR-handling strategies within the chained equations literature in \textbf{Supplementary Table B.1}. We implement the method in Tompsett et al. (2018) using the best possible value for the corresponding offset parameter related to $R_{.1}$.\cite{Tompsett2018} The method in Jolani (2012) implemented in \textit{mice.impute.ri} was also applied.\cite{Jolani2012} We found that the method in Jolani (2012) produced substantial residual bias in estimating outcome model parameters. The method in Tompsett et al. (2018) produced little bias when the corresponding offset sensitivity parameter was correctly specified. We observed similar results when these methods were applied to multiple variable missingness in \textbf{Supplementary Table C.1}.  \\
\indent In real data analysis, we will rarely know the true value of $\phi_1$ or any sensitivity parameter. We can apply these methods across different \textit{assumed} values for $\phi_1$ as a sensitivity analysis to departures from MAR. Results are shown in \textbf{Figure \ref{rezvan}b}. Point estimates from the method in Carpenter et al. (2007) did not vary much across assumed $\phi_1$ values above about 0.5.\cite{Carpenter2007} As demonstrated in \textbf{Supplementary Figure B.1}, this is because the true value of the outcome model parameter is outside the range of estimates obtained from MAR-based imputation (range between 0.32 and 0.45, mean = 0.39). In \textbf{Supplementary Figure B.2}, we compare the imputation-specific weights obtained for the method in Carpenter et al. (2007) and for our proposed method using the data visualization proposed in Heraud-Bousquet et al. (2012).\cite{Heraud-bousquet2012} The final weighted estimate for the Carpenter et al. (2007) method is dominated by the imputed dataset producing the most extreme MAR-based estimates for larger values of $\vert \phi_1 \vert$, with weights near 1 for a single imputed dataset.\cite{Carpenter2007} In contrast, the proposed method defines weights at the subject level, and the largest imputation-specific weight obtained for \textit{any} subject was near 0.5. For most subjects, the largest weight was about 0.06. These smaller imputation- and subject-specific weights produce a much more stable estimation of model parameters that is much less sensitive to extreme imputations drawn for individual subjects.  

\subsection{Estimation of standard errors}
\begin{figure}[htbp!]
  \centering
\caption{Coverage of 95\% confidence intervals for the coefficient of $Z_2$ from stacked and weighted analysis across 1000 simulated datasets assuming true $\phi_1$ is known}
\includegraphics[trim={0cm 0cm 0cm 1cm}, clip, width=6in]{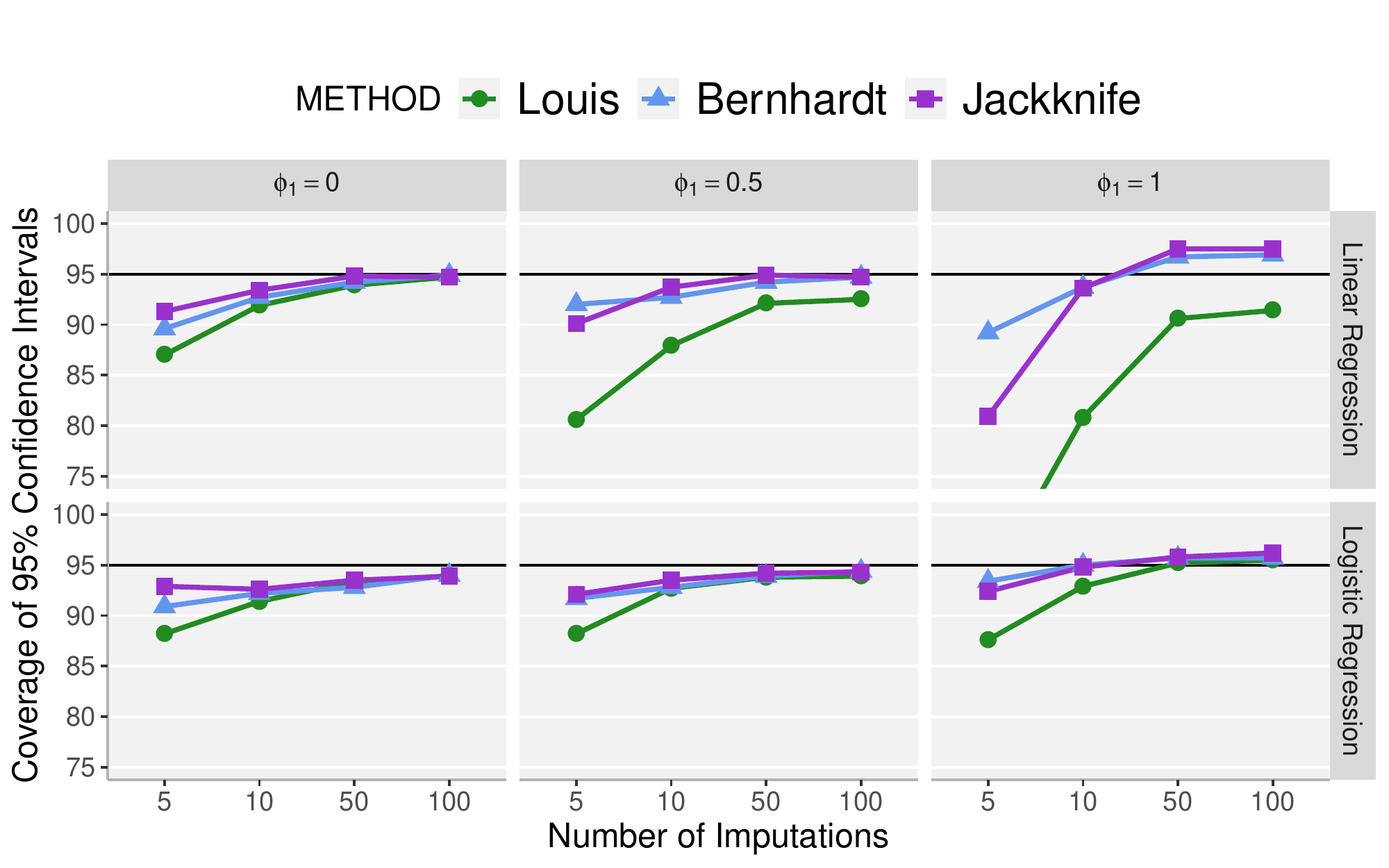}
\label{rezvancoverage}
\end{figure}

\indent We may also be interested in estimating standard errors for regression model parameters.  We apply the methods in \textbf{Section \ref{varestimation}} to estimate standard errors for the proposed stacked and weighted analysis method assuming that $\phi_1$ is correctly specified. We calculate corresponding 95\% confidence intervals for estimating the coefficient of $Z_2$ in linear or logistic regression models using each of the three variance estimation strategies and assuming a normal distribution approximation (point estimate $\pm$ 1.96*sqrt[standard error estimate]). Results are shown in \textbf{Figure \ref{rezvancoverage}}. When the target model was logistic regression, all of the variance estimation strategies produced similarly good coverage rates as long as $M$ is large enough (e.g., $>$50). However, the story is more complicated for coverage rates when the target outcome model is linear regression. When true $\phi_1$ is moderate or small (e.g., $\phi_1 \leq 0.5$), the method proposed in Bernhardt (2019) and our jackknife modification produced nominal coverage for large $M$.\cite{Bernhardt2019} However, these two approaches resulted in slight over-coverage when $\phi_1$ was large (e.g. 1). The method in \ref{louis} based on the observed data information principle in Louis (1982) resulted in small under-coverage for all values of $\phi_1 \neq 0$, with stronger under-coverage seen for larger $\phi_1$.\cite{Louis1982} \\
\indent As noted in Rezvan et al. (2015), the distribution actually used to impute normally-distributed $Z_{.1}$ within the chained equations imputation algorithm has heavier tails than the ``ideal" normal distribution when the corresponding dispersion parameter is not known.\cite{Rezvan2015} To evaluate whether this explains the observed over- and under-coverages, we repeated these simulations after imputing $Z_1$ from a normal distribution with dispersion parameter fixed at the simulation truth. This modified imputation strategy did not impact the resulting coverage rates (results not shown), indicating that this slight over- and under-coverage was not driven directly by the heavy-tailed imputation. Similar results were seen in simulations with multiple covariates as discussed in \textbf{Supplementary Section C}. \\
\indent \textbf{Supplementary Figure B.3} provides the average run-time for each of these variance estimation strategies under a normally-distributed imputed outcome with $n=1000$. At $M=100$, the method in \ref{louis} took on average only 1.4 seconds to estimate standard errors. In comparison, the method from Bernhardt (2019) and our jackknife modification took an average of roughly 39 and 25 seconds, respectively.\cite{Bernhardt2019} While this difference will be negligible for many analyses, the shorter runtime of the method in \ref{louis} may produce a much shorter aggregate runtime when sensitivity analyses are performed across a large grid of $\phi_1$ values.

\section{Illustrative example: missingness of HPV status in patients with oropharyngeal cancer}
\begin{figure}[htbp!]
  \centering
\caption{Estimated associations between HPV status and overall survival across assumed values for $\phi_1$ $^1$}
\includegraphics[trim={0cm 0cm 0cm 0.5cm}, clip, width=6in]{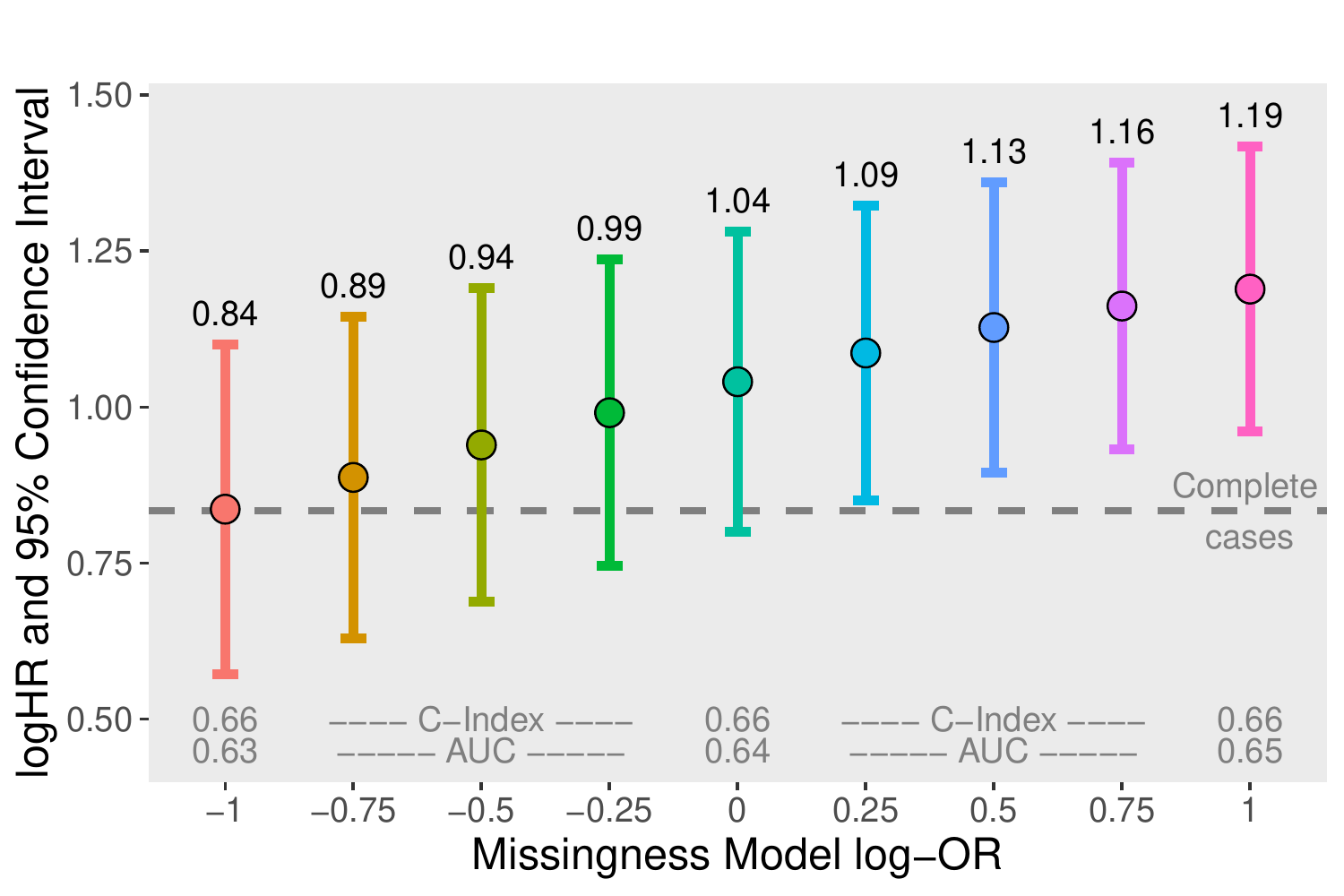}
\caption*{ \footnotesize $^1$ HR denotes hazard ratio, OR denotes odds ratio. For each value of $\phi_1$, point estimates come from a weighted Cox proportional model for overall survival adjusting for HPV status,  smoking status, ACE27 comorbidities, overall cancer stage, and age at cancer diagnosis. }
\label{sporeHPV}
\end{figure}

\indent We apply the proposed methods to address potential MNAR missingness in a study of N=840 patients newly-diagnosed with oropharyngeal squamous cell carcinoma at the University of Michigan. Baseline characteristics including smoking status, age, cancer stage, and comorbidities were collected at the time of study enrollment (at or soon after diagnosis), and patients were then followed for cancer-related outcomes including overall survival. For over 30\% of patients, HPV (human papillomavirus) status was not evaluated at baseline. Baseline comorbidities status (none/mild/moderate/severe) is missing for roughly 27\% of patients, and there is a very small amount of missingness in smoking status and cancer stage. For additional details about patient recruitment, data collection, and study descriptives, we refer the reader to Beesley et al. (2019).\cite{Beesley2019b}\\
\indent Suppose our interest lies in modeling overall survival as a function of baseline characteristics. For the baseline covariates of interest, only 45\% of patients have complete data. To improve estimation efficiency and guard against bias from complete case analysis, we want to perform multiple imputation to handle the missing data. However, we have concerns about the reasonableness of the MAR assumption. In particular, we consider missingness in HPV status. Missingness in HPV status is clearly related to year of enrollment for this cohort, reflecting the increasing acceptance of HPV status as a key prognostic factor for oropharyngeal cancer patients (\textbf{Supplementary Figure D.1}).  As shown in \textbf{Supplementary Table D.1}, missingness in HPV status is also associated with smoking status even after adjusting for year of enrollment, with current smokers being less likely to have HPV status available relative to never smokers (log-odds ratio -0.13, 95\% CI: -0.21, -0.04). In terms of oropharyngeal cancer etiology, this makes sense; two strong risk factors for oropharynx cancer development are smoking and HPV positivity. If the patient was a current smoker, it could be that doctors were less inclined to recommend testing for HPV status (at least, for earlier enrollment periods before HPV testing had become a standard part of patient care). However, additional unobserved factors such as sexual history could have also informed decision-making for whether a patient was tested for HPV infection. This could induce a MNAR association between HPV testing and true HPV status, where untested people may be less likely to be positive even after adjusting for other observed baseline variables and calendar time of enrollment. We are interested in exploring to what extent our estimated parameters in a model for overall survival are impacted by our assumptions regarding HPV missingness. \\
\indent We applied the method in \textbf{Figure \ref{methoddiagram}} as follows. First, we obtained 50 multiply imputed datasets assuming HPV missingness is MAR. Missingness for other baseline covariates was also assumed to be MAR. Details on this imputation procedure can be found in \textbf{Supplementary Section D}. We then obtained a stacked dataset and weighted the dataset proportional to $\text{exp}\left( -\phi_1 \mathcal{I}[\text{HPV positive}]\right)$, where $\phi_1$ corresponds to the log-odds ratio for \textit{observing} HPV status for HPV positive vs. HPV negative patients. We fit a weighted Cox proportional hazards regression model for overall survival using this stacked data, where $\phi_1$ was varied between -1 and 1. Relative to estimated log-odds ratios for missingness (\textbf{Supplementary Table D.1}), a value of $\vert \phi_1 \vert  > 0.2$ represents very extreme MNAR dependence. The Cox proportional hazards model adjusted for HPV status, smoking status, ACE27 comorbidities, overall cancer stage, and age at cancer diagnosis. We applied the method in \ref{louis} to estimate corresponding standard errors using the partial log-likelihood. \\
\indent \textbf{Figure \ref{sporeHPV}} shows the estimated log-hazard ratio for HPV positivity in the overall survival regression model as a function of $\phi_1$. We see that the point estimate for the log-hazard ratio does change in magnitude across $\phi_1$. For example, the log-hazard ratio is estimated as 1.19 (95\% CI: 0.96, 1.42) for $\phi_1 = 1$ and 0.84 (95\% CI: 0.57, 1.10) for $\phi_1 = -1$. However, for more reasonable values of $\vert \phi_1 \vert$ consistent with observed associations between missingness and other variables (\textbf{Supplementary Table D.1}), the log-hazard ratio only varies between about 0.99 and 1.09. Additionally, the 95\% confidence intervals across all $\phi_1$ are still far from 0 even for more extreme values of $\phi_1$. If our goal was to assess whether HPV status was associated with overall survival for patients newly-diagnosed with oropharyngeal cancer, our study findings would not be strongly impacted by assuming MAR. In this example, we may be more concerned about the magnitude of the association between HPV status and overall survival if we are wanting to use the resulting model to predict likely survival outcomes for new patients. To compare the impact of the choice of $\phi_1$ on discrimination of resulting 5-year overall survival estimates, we calculated 5-year survival predictions for each of the 378 patients with complete covariate data using each of the estimated survival models. We report the corresponding C-Indices and area under the receiver operating curve (AUC) in \textbf{Figure \ref{sporeHPV}}. The estimated C-Indices and AUC did not vary much across different values of $\phi_1$.

\FloatBarrier

\section{Discussion}
Multiple imputation by chained equations is an appealing approach to handling missing data in many data analysis settings. However, the majority of statistical development in this area relies on a key assumption that data are missing at random (MAR). Application of MAR-based imputation and data analysis strategies when data are not missing at random (MNAR) can produce bias in estimating parameters of interest.\cite{Little2002} \\
\indent In this paper, we propose a novel strategy for addressing single variable MNAR missingness \textit{given multiple imputations generated assuming MAR.} MNAR missingness is handled through weighted data analysis applied to the stacked multiple imputations, where the data weights are a function of an assumed model for MNAR missingness. In the special setting where the MNAR missingness mechanism can be reasonably approximated by a standard logistic regression, the weights take a simple form and depend only on a single sensitivity parameter. This parameter has a convenient interpretation as the log-odds ratio association between the MNAR missingness and the true value of the variable with MNAR missingness. \\
\indent The proposed method makes several advances over existing methods in this area. Unlike the related data re-weighting method in Carpenter et al. (2007), the proposed approach defines separate weights for each subject and imputation combination.\cite{Carpenter2007} This prevents estimation from being dominated by a single imputed dataset. As discussed in Rezvan et al. (2015), the method in Carpenter et al. (2007) may also produce inconsistent estimates of common parameters of interest (e.g. means, regression model parameters, etc.) in some cases.\cite{Rezvan2015} This is a result of the reliance on \textit{point estimates} obtained under MAR assumptions, which may be far from the truth. The proposed method uses \textit{imputed data} obtained under MAR assumptions but not the corresponding point estimates, avoiding this challenge and allowing for valid point estimation even under strong MNAR missingness.  \\
\indent Several authors have developed strategies for addressing MNAR missingness \textit{within} the chained equations imputation procedure itself. Tompsett et al. (2018) recommends including missing data indicators as predictors in the imputation model and handles MNAR missingness related to the imputed variable itself through a fixed offset with corresponding sensitivity parameter.\cite{Tompsett2018} This results in a regression model approximation of the ``exact" imputation distribution in \ref{imputeZ1}. This approach performed well in simulations when the corresponding offset parameter was well-specified (\textbf{Supplementary Tables B.1 and C.1}). However, imputation \textit{and} data analysis as in Tompsett et al. (2018) must be repeated across multiple values of the sensitivity parameter, which can become computationally challenging for a large grid of plausible values. Our proposed approach also involves repeated analysis across sensitivity parameter values, but it relies on a single set of multiple imputations, avoiding the need to re-impute the data many times. Jolani (2012) avoids use of sensitivity parameters entirely under assumptions that the true model generating missingness follows a logistic regression model with main effects.\cite{Jolani2012} However, we found that the implementation of this method in \textit{mice} in R performed poorly in terms of large residual bias in estimating regression model parameters. This may be related to difficulty in identifying parameters in the missingness model and warrants further exploration.\\
\indent One historical disadvantage of the general strategy of imputation stacking was the limited statistical literature regarding standard error estimation and the lack of corresponding software for easy implementation. However, Beesley and Taylor (2020) recently proposed a simple strategy for estimating standard errors for stacked and weighted multiple imputations (\ref{louis}) inspired by the observed data information principle in Louis (1982).\cite{Louis1982,Beesley2020} Bernhardt (2019) proposed an alternative strategy involving bootstrapping of multiply imputed data for estimating the between-imputation variation as in \ref{between}.\cite{Bernhardt2019} Through simulations in \textbf{Section \ref{rezvansimssec}}, we demonstrate that both general estimation strategies can produce reasonable standard error estimates, with some slight under-coverage (e.g. 90\%) seen for the method in \ref{louis} and some over-coverage seen for the method in Bernhardt (2019) when the MNAR missingness is strong. This under-coverage may be a result of the dependence between the weights (a function of the imputed data) and the target model parameter $\theta$ due to the model-based multiple imputation procedure, and future efforts can explore this issue in greater detail. An additional limitation of the proposed approach is that it generally requires more imputed datasets than Rubin's rules-based estimation. The number of required imputations will depend on the amount of missingness, but we generally found good estimation properties for $M=50$. Given sufficient $M$ and the correct sensitivity parameter, the proposed method resulted in very low bias even for small sample sizes (e.g. $n=100$).  \\
\indent Thus far, we have focused on the particular setting where we have MNAR missingness in a \textit{single} variable. Missingness in other variables was assumed to be MAR. In \textbf{Supplementary Section A}, we extend the proposed method to handle MNAR missingness in \textit{multiple} variables. In the special case where (1) missingness in each variable is independent of the true values for other variables with missingness and (2) the MNAR mechanisms are well-approximated with logistic regressions, resulting weights are similar to those in \ref{logistic} but with a separate sensitivity parameter for each variable that is MNAR. In simulations under (1) and (2), this extension performed as expected, with properties similar to those seen in simulations presented here (not shown). Future work can explore implementation for more general MNAR missingness. \\
\indent An overall advantage of the proposed method is that it disentangles the challenges of data imputation (i.e., filling in the missing values) and handling of MNAR (i.e., avoiding or reducing bias due to the MNAR missingness mechanism). This approach can be applied to data previously imputed under MAR assumptions, and point estimation can be very easily implemented using standard software. Standard error estimation presents a greater challenge, and we provide R package \textit{StackImpute} (available at \url{https://github.com/lbeesleyBIOSTAT/StackImpute}) to allow users to easily obtain standard errors for many commonly-used regression model settings, including Cox proportional hazards regression and generalized linear models.

\section*{Acknowledgments}
\indent The authors cite the many investigators in the University of Michigan Head and Neck Specialized Program of Rese`arch Excellence for their contributions to patient recruitment, specimen collection, and study conduct. This research is partially supported by National Institutes of Health grant CA129102.

\section*{Data Availability}
Data from illustrative example are not shared due to third-party data sharing restrictions and to protect patient privacy.

\bibliographystyle{plainnat}
\bibliography{Bib}%

\end{document}